\documentclass[aps,prb,preprint,showpacs,preprintnumbers,amsmath,amssymb]{revtex4}
\usepackage{amsmath}
\usepackage{amssymb}
\usepackage{amsfonts}
\usepackage{color,array}
\usepackage{dsfont}
\usepackage{slashed}
\usepackage{graphicx}
\usepackage{graphics}
\usepackage{epsfig}
\usepackage{bbm,bm}
\usepackage{psfrag}
\usepackage{ulem}
\usepackage{url}

\begin{document}

\title{Critical Exponents of Superfluid Helium and
Pseudo-$\epsilon$ Expansion}

\author{A. I. Sokolov}
\email{ais2002@mail.ru}
\author{M. A. Nikitina}
\affiliation{Department of Quantum Mechanics,
Saint Petersburg State University,
Ulyanovskaya 1, Petergof,
Saint Petersburg, 198504
Russia}
\date{\today}

\begin{abstract}
Pseudo-$\epsilon$ expansions ($\tau$-series) for critical exponents of 3D XY
model describing $\lambda$-transition in liquid helium are derived up to
$\tau^6$ terms. Numerical estimates extracted from the $\tau$-series obtained
using Pad\'e-Borel resummation technique, scaling relations and seven-loop
($\tau^7$) estimate for the Fisher exponent $\eta$ are presented including those
for exponents $\alpha$ and $\nu$ measured in experiments with record accuracy.
For the exponent $\alpha$ the procedure argued to be most reliable gives
$\alpha= -0.0117$. This number is very close to the most accurate experimental
values differing appreciably from the results of numerous lattice and
field-theoretical calculations. It signals that the pseudo-$\epsilon$ expansion
is a powerful tool robust enough to evaluate critical exponents with very small
absolute error. The arguments in favour of such a robustness are presented.
\end{abstract}


\maketitle

\section{Introduction}

Nowadays there exists a great number of high-precision numerical estimates
of critical exponents and other universal quantities for three-dimensional
systems obtained within various theoretical approaches. High-temperature
expansion technique, Monte Carlo simulations, field-theoretical
renormalization-group analysis based upon many-loop calculations in three
and ($4 - \epsilon$) dimensions are among them (see, e. g. Refs.\cite{ZJ,
Kl2001,PV02}). In most cases, an agreement between theoretical estimates is
so good and their (apparent) accuracy is so high that experimental results
being, as a rule, less accurate start to lose their fundamental role in the
physics of critical phenomena. Phrase "Experimentalists can only confuse
us" said at the International Workshop in Bad Honnef\cite{BH95} sounds today
even more actual than 20 years ago.

At the same time, there is an area within the phase transition science
where experiment certainly passes ahead a theory. We mean the physics of
superfluid transition in liquid helium-4. Traditionally\cite{FBK58,L92,GMA93},
experimental study of thermodynamic and kinetic properties of this quantum
fluid is carried out on the very high technical level and covers the
temperatures extremely close to the $\lambda$ point. This, in particular,
enabled to measure critical exponents of superfluid helium with
unprecedented accuracy, including the specific heat exponent $\alpha$
which is known to be tiny. Record measurements \cite{L96,L2000,L03}
performed in space in order to avoid the influence of gravity yielded
$\alpha = -0.0127\pm 0.0003$, the value of critical exponent that is
accepted as the most accurate ever obtained experimentally.

In general, most of theoretical data agree or almost agree with the results of
experimental determination of critical exponents for superfluid helium provided
the uncertainty of computations is estimated in conservative enough way.
Lattice estimates of $\alpha$ exhibit a tendency to group around
$-0.015$\cite{BC99,HT99,CHPRV01,CHPV06,BMPS06} while their field-theoretical
counterparts lie mainly between $-0.004$ and $-0.013$\cite{LGZ77,BNM78,LGZ80,
AS95,GZJ98,Kl99,Kl2000,Kl01,PS07}, i. e. oppositely regarding the experimental
value mentioned. This discrepancy being small is nevertheless attracts
attention (see, e. g. Refs.\cite{CHPRV01,CHPV06,PS07}) and ways to resolve it
are permanently looked for.

In such a situation it is resonable to evaluate the critical exponents for
superfluid transition in helium-4 within an alternative approach which proved to
be highly efficient numerically in the phase transition problem. We mean the
method of pseudo-$\epsilon$ expansion invented by B. Nickel many years ago
(see Ref. 19 in the paper of Le Guillou and Zinn-Justin \cite{LGZ80}). This method
was applied to various systems\cite{FHY2000,COPS04,HID04,CP05,S2005,NS13,NS14c}
including two dimensional and those with non-trivial symmetry of the order
parameter and lead to rather good numerical results for all the models considered.
High numerical power of pseudo-$\epsilon$ expansion technique stems from its key
feature: it transforms strongly divergent renormalization group (RG) expansions
into the series having smaller lower-order coefficients and much slower growing
higher-order ones what makes them very convenient for getting numerical estimates.
Moreover, as was recently shown\cite{NS14e}, the pseudo-$\epsilon$ expansion
machinery works well even in the case of the Fisher exponent $\eta$ when original
RG expansion has irregular structure and is quite unsuitable for extracting
numerical results.

Below, the pseudo-$\epsilon$ expansions ($\tau$-series) for critical exponents of
three-dimensional XY model will be calculated starting from the six-loop\cite{BNM78}
RG series. The $\tau$-series for the exponents $\alpha$ and $\gamma$ will be written
down up to $\tau^6$ terms. Numerical estimates for the critical exponents will be
obtained using Pad\'e-Borel resummation technique, scaling relations and the
seven-loop ($\tau^7$) pseudo-$\epsilon$ expansion estimate for Fisher exponent
$\eta$. Comparing the numbers obtained with the results of the most advanced
experiments and with the values extracted from lattice and field-theoretical
calculations the numerical effectiveness of the pseudo-$\epsilon$ expansion approach
will be evaluated. The general properties of this approach will be discussed and
the roots of its high numerical power will be cleared up.

\section{Pseudo-$\epsilon$ expansions for critical exponents $\alpha$ and $\gamma$.
Numerical estimates}

Critical thermodynamics of three-dimensional XY model is described by Euclidean
field theory with the Hamiltonian:
\begin{equation}
\label{eq:1}
H = \int d^{3}x \Biggl[{1 \over 2}( m_0^2 \varphi_{\alpha}^2
 + (\nabla \varphi_{\alpha})^2)
+ {\lambda \over 24} (\varphi_{\alpha}^2)^2 \Biggr] ,
\end{equation}
where $\alpha = 1, 2$, bare mass squared $m_0^2$ is proportional to $T -
T_c^{(0)}$, $T_c^{(0)}$ being the mean field transition temperature.
Perturbative expansions for the $\beta$-function and critical exponents
for the model (1) have been calculated within the massive theory
\cite{BNM78,MN91} with the propagator, quartic vertex and $\varphi^2$
insertion normalized in a standart way:
\begin{eqnarray}
\label{eq:4}
G_R^{-1} (0, m, g_4) = m^2 , \qquad \quad {{\partial G_R^{-1}
(p, m, g_4)} \over {\partial p^2}}
\bigg\arrowvert_{p^2 = 0} = 1 , \\
\nonumber
\Gamma_R (0, 0, 0, m, g) = m^2 g_4, \qquad \quad
\Gamma_R^{1,2} (0, 0, m, g_4) = 1.
\end{eqnarray}

We derive pseudo-$\epsilon$ expansions ($\tau$-series) for critical
exponents $\alpha$ and $\gamma$ starting from corresponding six-loop RG
series. To find these pseudo-$\epsilon$ expansions one has to substitute
$\tau$-series for Wilson fixed point coordinate $g^*$ into perturbative RG
series for critical exponents and reexpand them in $\tau$. With
$\tau$-series for $g^*$\cite{NS14c} and RG expansions of exponents in hand
the calculations are straightforward. Their results read:
\begin{eqnarray}
\alpha = {1 \over 2} &-& {3\tau \over 10} - 0.1297777778 \tau^2
- 0.039547352 \tau^3 - 0.02432025 \tau^4
\nonumber\\
&-& 0.00324983 \tau^5 - 0.0121092 \tau^6.
\end{eqnarray}
\begin{eqnarray}
\gamma^{-1} = 1 &-& \frac{\tau}{5} - 0.0405925926 \tau^{2} +
0.004326858 \tau^{3} - 0.00566467 \tau^{4}
\nonumber\\
&+& 0.00458218 \tau^{5} - 0.0067372 \tau^{6}.
\end{eqnarray}
We present here the pseudo-$\epsilon$ expansion for inverse $\gamma$ instead
of $\tau$-series for the exponent $\gamma$ itself because the former turns
out to be more suitable for getting numerical estimates. We do not present
$\tau$-series for critical exponent $\nu$ since it can be easily deduced
from (3) using well known scaling relation
\begin{eqnarray}
\alpha = 2 - D\nu.
\end{eqnarray}

Despite of small and rapidly decreasing coefficients pseudo-$\epsilon$
expansions (3), (4) remain divergent. So, to extract numerical values of
critical exponents from these series one has to apply some resummation
procedure. We employ Pad\'e-Borel resummation technique which is based on
the Borel transformation
\begin{equation}
f(x) = \sum_{i = 0}^{\infty} c_i x^i = \int\limits_0^{\infty} e^{-t}
F(xt) dt , \qquad F(y) = \sum_{i = 0}^{\infty} {\frac{c_i }{i!}} y^i .
\end{equation}
and use of Pad\'e approximants [L/M] for analytical continuation of the
Borel transform $F(y)$. Application of this technique to $\tau$-series for
$\alpha$ leads, however, to the results which are far from to be
satisfactory. This is seen from Table I representing Pad\'e-Borel triangle
for the series (3). More than a half of nontrivial estimates are absent in
this table because of positive axis ("dangerous") poles spoiling
corresponding Pad\'e approximants. Existing estimates are strongly
scattered being practically useless for getting accurate value of the
exponent $\alpha$. Moreover, even more conservative procedure that uses
simple Pad\'e approximants and gives numerical results much less sensitive
to the problem of poles results in numbers appreciably differing from each
other even in the highest ($\tau^6$) order available. Table II
representing Pad\'e triangle for the series (3) demonstrates this fact.

In such a situation it is natural to evaluate the exponent $\alpha$ in a
different manner, using the scaling relation containing critical exponents
$\gamma$ and $\eta$. It is readily obtained combining (5) with
\begin{eqnarray}
\nu = {\gamma \over {2 - \eta}}.
\end{eqnarray}
This way to evaluate $\alpha$ looks attractive because of two reasons.
First, Pad\'e-Borel estimates of $\gamma$ resulting from $\tau$-series (4)
converge to the asymptotic value very rapidly signaling that for this
exponent the iteration procedure employed is rather efficient. This is
clearly seen from Table III where the Pad\'e-Borel triangle for the
exponent $\gamma$ is presented. Second, the numerical value of the Fisher
exponent can be extracted from the recently found seven-loop
$\tau$-series\cite{NS14e}, i. e. it can be obtained with the highest
accuracy accessible within the pseudo-$\epsilon$ expansion approach.

As is well known, diagonal and near-diagonal Pad\'e approximants possess
the best approximating properties\cite{BGM}. That is why the value
$\gamma = 1.3156$ given by approximant [3/3] (see Table III) is assumed
to be the most reliable one resulting from the $\tau$-series for
$\gamma^{-1}$. Its counterpart originating from the $\tau$-series for
$\gamma$ itself resummed within Pad\'e-Borel technique using approximant
[3/3] is equal to 1.3162. So, the average over these two numbers will be
accepted as a final pseudo-$\epsilon$ expansion estimate for the
susceptibility critical exponent: $\gamma = 1.3159$. The pseudo-$\epsilon$
expansion estimate for $\eta$ is extracted from seven-loop
$\tau$-series\cite{NS14e}
\begin{eqnarray}
\eta &=& 0.0118518519 \tau^2 + 0.0105390747 \tau^3 + 0.005188190 \tau^4
\nonumber\\
&+& 0.003229563 \tau^5 + 0.00145159 \tau^6 + 0.0016264 \tau^7
\end{eqnarray}
by means of the same, Pad\'e-Borel resummation procedure. Use of
near-diagonal Pad\'e approximant [4/3] free of dangerous poles leads to
$\eta = 0.0376$. This value agrees well with the results of alternative
field-theoretical calculations $\eta = 0.0380(50)$ ($\epsilon$-expansion),
$\eta = 0.0370(50)$ (biased $\epsilon$-expansion), and $\eta = 0.0354(25)$
(3D RG)\cite{GZJ98} and is accepted as a pseudo-$\epsilon$ expansion
estimate for the Fisher exponent.

It is worthy to evaluate the accuracy of numerical results thus found.
To do this we adopt the following strategy. We assume that the difference
between the numbers obtained from the same $\tau$-series by means of various
resummation procedures is a natural measure of numerical accuracy provided
by the pseudo-$\epsilon$ expansion approach. In the case of critical
exponents $\gamma$, use of simple Pad\'e approximants to resum the series
(4) gives the ultimate value 1.3154 deviating from above Pad\'e-Borel
estimate by 0.0005. This value is accepted to be a characteristic error of
our estimate for $\gamma$. For the Fisher exponent Pad\'e estimate
originating from the seven-loop $\tau$-series is equal to 0.0348\cite{NS14e}.
Hence, the characteristic error in this case equals to 0.0028. For the
exponent $\alpha$ evaluated via $\gamma$ and $\eta$ with a help of formulas
(5) and (7) it leads to $\Delta\alpha=0.0029$.

Using known scaling relations and estimating the accuracy of numerical
results in the way just described we arrive to the following set of critical
exponents for the $\lambda$-transition in liquid helium-4:
\begin{equation}
\alpha = -0.0117(29), \quad \nu = 0.6706(12), \quad \gamma = 1.3159(5),
\quad \eta = 0.0376(28), \quad \beta = 0.3479(15).
\end{equation}
Let us compare these values with experimental data and with the numbers
extracted from field-theoretical and lattice calculations. Since critical
exponent $\alpha$ is what is measured in experiments with highest
accuracy\cite{L03} along with the correlation length exponent\cite{GMA93}
related to $\alpha$ by the scaling relation (5) we concentrate here on the
data for $\alpha$. They are collected in Table IV. As is seen, our
pseudo-$\epsilon$ expansion estimate is in a good agreement with the
experimental data but deviates appreciably from the most of the results of
RG analysis in three and ($4-\epsilon$) dimensions and from the lattice
estimates. Hence, addressing the pseudo-$\epsilon$ expansion approach
enables one to avoid discrepancy between theoretical predictions and the
results of most accurate measurements.

\section{Pseudo-$\epsilon$ expansion machinery is robust}

So, the pseudo-$\epsilon$ expansion approach results in iterations that
converge to the high-precision values of critical exponents. It
demonstrates a robustness of this approach that may be argued to be its
general property. Indeed, let the pseudo-$\epsilon$ expansion for the
Wilson fixed point location $g^*$ be:
\begin{equation}
g^* = \tau + A \tau^2 + B \tau^3 + C \tau^4 + D \tau^5 + ... ,
\end{equation}
while field-theoretical RG series for some critical exponent $\zeta$ have a form:
\begin{equation}
\zeta = p_0 + p_1 g + p_2 g^2 + p_3 g^3 + p_4 g^4 + p_5 g^5 + ... .
\end{equation}
Then, to obtain $\tau$-series for $\zeta$, we have to substitute expansion
(10) into (11). It yields:
\begin{equation}
\zeta = p_0 + p_1 \tau + (A p_1 + p_2) \tau^2 + (B p_1 + 2A p_2 + p_3) \tau^3
+ [C p_1 + (A + 2B) p_2 + 3A p_3 + p_4)] \tau^4 + ...
\end{equation}

As seen from (12) the coefficient of $k$-th term in the pseudo-$\epsilon$
expansion for $\zeta$ depends not only on the coefficients of the same
order in series (10) and (11) but is determined by all the coefficients of
$k$-th and lower orders starting from $A$ and $p_1$. It means that
applying pseudo-$\epsilon$ expansion approach one uses the information
contained in the known terms of original RG expansions to a greater extent
than when conventional resummation procedures employed. This point is
essential since the known coefficients of perturbative RG expansions are,
in fact, the only input data we really use to evaluate the critical
exponents and other universal quantities. All the rest information, e. g.
the character of asymptotic behavior of coefficients under $k \to \infty$
is employed to choose the resummation procedure assumed to be optimal, to
evaluate the (apparent) accuracy of numerical values obtained, etc., but
not to fix the numbers themselves. Pseudo-$\epsilon$ expansion machinery
realizing multiple use of the results of perturbative RG calculations is
in this sense more robust than other resummation methods. High numerical
efficiency of the pseudo-$\epsilon$ expansion approach demonstrated above
may be referred to as manifestation of this robustness.

It is interesting that our estimate of $\alpha$ turns out to be close to
that given by direct summation of corresponding $\tau$-series ($-$0.0090).
Although such a closeness may be thought of as occasional, it looks
symptomatic and confirms the conclusion that pseudo-$\epsilon$ expansion
is a robust procedure.

What would shed more light on the general properties and numerical power
of the approach discussed is the knowledge of large-order behavior of
pseudo-$\epsilon$ expansion coefficients. Today such information is absent.
We believe, however, that it will be obtained in near future.

\section{Conclusion}

To summarize, we have calculated pseudo-$\epsilon$ expansions for critical
exponents of the three dimensional XY model up to $\tau^6$ order. Numerical
estimates have been found by means of Pad\'e-Borel resummation of
$\tau$-series for the exponents $\gamma$ combined with a use of scaling
relations and numerical value of the Fisher exponent extracted from
Pad\'e-Borel resummed seven-loop $\tau$-series. The values of critical
exponents $\alpha$ and $\nu$ thus obtained turn out to be in a good agreement
with the data of most precise measurements including those performed in space.
It has been argued that pseudo-$\epsilon$ expansion approach represents
resummation procedure that exploits the information contained in known
coefficients of RG series to a greater extent than conventional resummation
methods do. This may be the origin of its high numerical effectiveness that
manifests itself, in particular, when critical exponents as tiny as the
exponent $\alpha$ for superfluid helium are evaluated.

\begin{table}[t]
\caption{Pad\'e-Borel table for pseudo-$\epsilon$ expansion of critical
exponents $\alpha$. Many estimates are absent because corresponding Pad\'e
approximants are spoilt by dangerous (positive axis) poles.}
\label{tab1}
\renewcommand{\tabcolsep}{0.4cm}
\begin{tabular}{{c}|*{8}{c}}
$M \setminus L$ & 0 & 1 & 2 & 3 & 4 & 5 & 6 \\
\hline
0 & 0.5000 & 0.2000 & 0.0702    & 0.0307    & 0.0064 & 0.0031 & $-$0.0090 \\
1 & 0.3456 &   $-$  &   $-$     &   $-$     &   $-$  &   $-$  &           \\
2 & 0.2927 & 0.0526 &   $-$     &   $-$     &   $-$  &        &           \\
3 & 0.2693 &   $-$  & $-$0.0032 & $-$0.0063 &        &        &           \\
4 & 0.2570 & 0.0534 &   $-$     &           &        &        &           \\
5 & 0.2498 &   $-$  &           &           &        &        &           \\
6 & 0.2452 &        &           &           &        &        &           \\
\end{tabular}
\end{table}

\begin{table}[t]
\caption{Pad\'e triangle for the pseudo-$\epsilon$ expansion of critical exponents
$\alpha$.}
\label{tab2}
\renewcommand{\tabcolsep}{0.4cm}
\begin{tabular}{{c}|*{8}{c}}
$M \setminus L$ & 0 & 1 & 2 & 3 & 4 & 5 & 6 \\
\hline
0 & 0.5000 &    0.2000 &    0.0702 &    0.0307 &    0.0064 & 0.0031 & $-$0.0090 \\
1 & 0.3125 & $-$0.0287 &    0.0133 & $-$0.0325 &    0.0026 & 0.0075 &            \\
2 & 0.2253 &    0.0161 & $-$0.0025 & $-$0.0035 & $-$0.0082 &        &            \\
3 & 0.1769 & $-$0.0212 & $-$0.0035 & $-$0.0023 &           &        &            \\
4 & 0.1451 &    0.0077 & $-$0.0077 &           &           &        &            \\
5 & 0.1231 & $-$0.0354 &           &           &           &        &            \\
6 & 0.1064 &           &           &           &           &        &            \\
\end{tabular}
\end{table}

\begin{table}[t]
\caption{Pad\'e-Borel table for the critical exponent $\gamma$ obtained
from pseudo-$\epsilon$ expansion for $\gamma^{-1}$. Several estimates are
absent because corresponding Pad\'e approximants have poles on the positive
real axis. The lowest line (RoC) demonstrates the rate of convergence of
Pad\'e-Borel estimates for $\gamma$ to the asymptotic value. Here the
estimate of $k$-th order is that given by diagonal approximant or by average
over two near-diagonal ones when corresponding diagonal approximant does
not exist.}
\label{tab3}
\renewcommand{\tabcolsep}{0.4cm}
\begin{tabular}{{c}|*{8}{c}}
$M \setminus L$ & 0 & 1 & 2 & 3 & 4 & 5 & 6 \\
\hline
0 & 1       & 1.2500 & 1.3168 & 1.3094 & 1.3191 & 1.3112 & 1.3229 \\
1 & 1.1736  &   $-$  & 1.3103 & 1.3133 & 1.3150 & 1.3157 &        \\
2 & 1.2363  & 1.3013 & 1.3159 & 1.3153 &   $-$  &        &        \\
3 & 1.2609  &   $-$  & 1.3152 & 1.3156 &        &        &        \\
4 & 1.2738  & 1.3047 &   $-$  &        &        &        &        \\
5 & 1.2908  &   $-$  &        &        &        &        &        \\
6 & 1.2853  &        &        &        &        &        &        \\
\hline
RoC & 1     & 1.2118 & 1.2766 & 1.3058 & 1.3159 & 1.3152 & 1.3156 \\
\end{tabular}
\end{table}

\begin{table}[t]
\caption{The values of critical exponent $\alpha$ obtained in this work, found
in experiments and extracted from resummed 3D RG series, $\epsilon$-expansions
and lattice calculations (LC).}
\label{tab4}
\renewcommand{\tabcolsep}{0.4cm}
\begin{tabular}{|c|cccc|}
\hline
This work & $-0.0117(29)$ &  &  &  \\
\hline
Experiments & $-0.0127(3)$\cite{L03} & $-0.0115(18)$\cite{GMA93}
& $-0.0124(12)$\cite{L92} &   \\
\hline
3D RG  & $-0.008(3)$\cite{LGZ77}  & $-0.007(9)$\cite{BNM78} & $-0.007(6)$\cite{LGZ80}
& $-0.010$\cite{AS95}  \\
& $-0.011(4)$\cite{GZJ98} & $-0.01294(60)$\cite{Kl99} & $-0.01126(100)$\cite{Kl2000}
& $-0.0100(20)$\cite{PS07}  \\
\hline
$\epsilon$-expansion & $-0.004(1)$\cite{GZJ98} & $-0.013$\cite{GZJ98}(biased)
& $-0.0091(39)$\cite{Kl01} &    \\
\hline
LC & $-0.0169(33)$\cite{HT99} & $-0.0146(8)$\cite{CHPRV01} & $-0.0151(3)$\cite{CHPV06}
& $-0.0151(9)$\cite{BMPS06}  \\
\hline
\end{tabular}
\end{table}

\end{document}